\documentstyle[aaspp4m,epsf,11pt]{article}
\def\jvec#1{\vec{\bf #1}}
\newcommand{\figcomment}[1]{#1}
\def\unsetyr{\def\oyear{\relax}\def\cyear{\relax}\def\cyeara{a\relax}\def\cyearb{b\relax}\def\cyearc{c\relax}\def\cyeard{d\relax}}
\def\setyr{\def\oyear{(}\def\cyear{)}\def\cyeara{a)}\def\cyearb{b)}\def\cyearc{c)}\def\cyeard{d)}}
\unsetyr
\def\jcite#1{\setyr\cite{#1}\unsetyr}

\def\rmmat#1{{\hbox{\rm #1}}}
\def\rmscr#1{\rmmat{\scriptsize #1}}
\newcommand{\be}{\begin{equation}}
\newcommand{\ee}{\end{equation}}
\newcommand{\bt}{\begin{table} \begin{center}}
\newcommand{\et}{\end{center} \end{table}}
\newcommand{\ba}{\begin{eqnarray}}
\newcommand{\ea}{\end{eqnarray}}
\newcommand{\ie}{{\it i.e.~}}
\newcommand{\eg}{{\it e.g.~}}
\newcommand{\cf}{{\it c.f.~}}
\def\d{{\rm d}}

\def\dd#1#2{\frac{\d #1}{\d #2}}

\newcommand{\comment}[1]{\relax}
\def\eqref#1{Equation~\ref{eq:#1}}

\def\figref#1{Figure~\ref{fig:#1}}

\def\rcwx{1E~161348-5055}

\begin{document}
\title{Hotspot Emission from a Freely Precessing Neutron Star}
\author{Jeremy S. Heyl}
\authoremail{jsheyl@tapir.caltech.edu}
\affil{
Theoretical Astrophysics,
mail code 130-33,
California Institute of Technology,
Pasadena CA 91125}
\and
 \author{Lars Hernquist}
\authoremail{lars@kona.harvard.edu}
\affil{
Harvard-Smithsonian Center for Astrophysics, 
60 Garden Street
Cambridge, MA 02138}

\begin{abstract}
Recent observations of \rcwx\, the neutron-star candidate at the center
of the supernova remnant RCW 103, show that a component of its emission
varies sinusoidally with a period of approximately six hours.  We
argue that this period is what one would expect for a freely
precessing neutron star with a spin period of about one second.  We
produce light curves for a freely precessing neutron star with a
hotspot.  By a suitable choice of parameters, we obtain light curves
which are constant with rotational phase when the flux from the star
reaches a maximum.  At other phases of the precession, the flux varies
as the star rotates but the total flux decreases by a factor of
several.  These models can explain the behavior observed from \rcwx\
and predict that the spin period should be detectable at minimum
flux from sufficiently sensitive measurements.
\end{abstract}

\section{Introduction}

\jcite{Garm00} have noted that the x-ray source, \rcwx , located near
the center of the supernova remnant RCW 103, has a sinusoidal light
curve with a period of approximately six hours.  The authors suggest
that this period implies that the x-ray source has a low-mass companion
with a six hour orbital period.  In this paper, we examine
an alternative possibility, that \rcwx\ is a freely precessing neutron
star with a precession period of about six hours.

The varying spectrum from \rcwx\ is similar to a blackbody yielding
an effective area of a small fraction of a square kilometer (Garmire,
private communication), much less than the surface area of a
neutron star; therefore, the situation could be well approximated by a
point source (a hotspot) on the surface of a freely-precessing
neutron star.

Free precession has often been invoked to explain long-period
variability in neutron stars and neutron-star systems.
\jcite{1975Natur.257..203B} attributed the 35-day cycle of Her X-1
(\cite{1972ApJ...174L.143T}) to the free precession of the neutron
star secondary.  \jcite{1988MNRAS.235..545J} explained timing
residuals in the Crab pulsar as arising
from small amplitude free precession of the
neutron star. \jcite{1996A&A...306..443C} found a small amplitude
modulation in the optical flux from the Crab pulsar which
\jcite{1997A&A...324.1005C} cite as evidence of free precession.

In \S \ref{sec:freep} we outline the kinematics of free precession,
and in \S \ref{sec:lensing}, we review the equations that determine
the trajectory of light leaving the surface of the neutron star.
\S \ref{sec:lightc} presents the light curves as a function of
precessional and rotational phase for Newtonian, relativistic and
ultracompact neutron stars with a emission from a hotspot.  Finally,
\S \ref{sec:disc} places the results in a greater context.

\section{Free Precession}
\label{sec:freep}
A body will precess freely if its rotation axis does not coincide with
one of its principal axes.  More specifically, if the body is only
slightly prolate or oblate, the angular velocity vector 
($\jvec{\omega}$) of the star will make a
constant angle ($\kappa$) with one of the principal axes (the $3-$axis),
forming the body cone, and will trace a cone in space (the space cone) with
half-opening angle $\kappa$ as well.   The rate of the precession is given
by (\cite{Gold80})
\be
\Omega = \frac{I_3-I_1}{I_1} \omega_3 = \epsilon \omega_3.
\ee
Values of $\epsilon=10^{-3} - 10^{-4}$ agree with the glitching
behavior of neutron stars and the inferred gravitational-radiation
spindown from the Crab pulsar (\cite{Shap83}).  Forced precession by
an orbiting companion typically has a frequency lower by a factor of
$2/ (3 \Omega_* \omega_3)$ where $\Omega_*$ is the angular frequency of
the orbit.

Internal magnetic fields may distort a neutron star significantly.
\jcite{1969ApJ...157.1395O} estimate the distortion of a neutron star
due to internal fields,
\be
\epsilon \simeq 4 \times 10^{-6} \left ( 3 \left < B_{p,15}^2 \right >
- \left < B_{\phi,15}^2 \right > \right )
\label{eq:bdistort}
\ee
where $B_{15}$ is the value of the magnetic field in units of
$10^{15}$~G and $\left < \ldots \right >$ denotes a volume-weighted average
over the star.  Here, $B_p$ and $B_\phi$ refer to the poloidal and
azimuthal components of the magnetic field, respectively.
One expects the internal fields of the neutron star to
be significantly larger than the field inferred by 
magnetic dipole radiation
due to the contribution of higher multipoles and the concentration of
magnetic flux in flux tubes (\eg \cite{1985Natur.316...27P}).

We would like to calculate the light curve from a hotspot on the
surface of the freely precessing neutron star.  Let us take the center
of the cone that the angular velocity moves along to be the
$\jvec{z}-$axis. Our line of sight ($\jvec{O}$) makes an angle $\xi$ with
this axis and forms a plane with $\jvec{z}-$axis.  We measure the
phase of the precession ($\phi$) relative to intersection of the space
cone with this plane.

The plane containing $\jvec{z}$ and $\jvec{O}$ also intersects the
body cone.  The angular momentum of the star points along the
$\vec{\bf z}$-axis.  In the body frame, the angular velocity of the
star traces a cone centered on a principal axis of the star.  We call
this principal axis $\jvec{3}$.  The hotspot is located at
$\jvec{\mu}$ which makes an angle $\beta$ with $\jvec{3}$.  Let us
freeze the precession and the rotation of the star when $\jvec{3}$
points along $\jvec{z}$ (see \figref{geom}).  At this orientation, the
angle between the $\jvec{3}-\jvec{O}$ plane and the
$\jvec{3}-\jvec{\mu}$ plane is $\gamma$.

Using spherical trigonometry (see \figref{geom})
yields the angle between the line of
sight and the rotation axis, $\zeta(\phi)$, and the angle between the
hotspot and the rotation axis $\alpha(\phi)$.
\ba
\cos\zeta(\phi) &=& \cos\kappa \cos\xi + \sin\kappa \sin\xi \cos \phi \\
\cos\alpha(\phi) &=& \cos\kappa \cos\beta + \sin\kappa \sin\beta \cos(\phi-\gamma) 
\ea
\figcomment{
\begin{figure}
\plotone{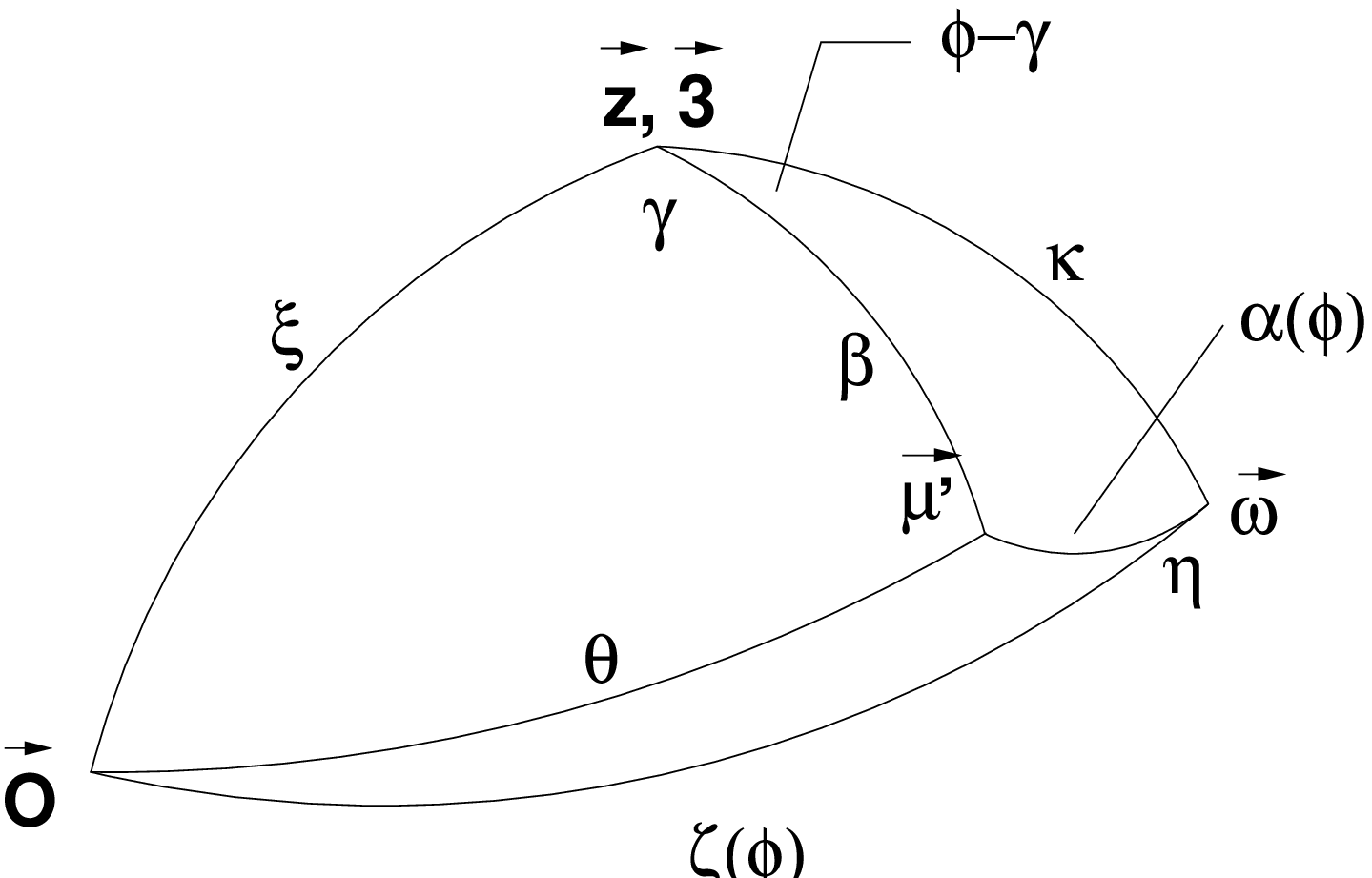}
\caption{Definitions of various angles}
\label{fig:geom}
\end{figure}
}

The star rotates as well as precesses.  The phase of the rotation is
given by $\eta$.  When $\eta=0$ the hotspot lies in the
$\jvec{\omega}-\jvec{O}$ plane.  The angle between the line of sight
and the hotspot is $\theta$ and is given by
\be
\cos \theta = \cos\alpha(\phi) \cos\zeta(\phi) +
\sin\alpha(\phi) \sin\zeta(\phi) \cos\eta
\label{eq:costheta}
\ee
If the hotspot emits isotropically, and we neglect gravitational
lensing, the observed flux from the hotspot is simply proportional to
$\cos \theta$ for $|\theta| < \pi/2$ and zero otherwise.

\section{Gravitational Lensing}
\label{sec:lensing}

\jcite{Page95} presents a detailed treatment of the gravitational
lensing of the surface of a neutron star.  Since the light trajectory 
is bent, the zenith angle of our detector ($\delta$) as seen from the 
hotspot is no longer equal to $\theta$.  They are related by
\be
\theta(x) = \int_0^y { x d u \over \sqrt{(1-2y) y - (1- 2 u) u^2 x^2}}
\ee
where $y=G M / R c^2$ and $x=\sin\delta$.  For $y<1/3$, the image of
the hotspot $i$ will be visible if $\theta_i+2\pi j<\theta(1)$.  The
flux from the hotspot is proportional to
\be
\sum_j \frac{x(\theta_i+2\pi j)}{\sin(\theta_i+2\pi j)}
\left. \dd{x}{\theta} \right|_{\theta=\theta_i+2\pi j}
\ee
such that $|\theta_i+2\pi j| \leq \theta(1)$.

For $y>1/3$ the surface of the neutron star lies below the circular
photon orbit; therefore, each hotspot yields an infinite number of
images (\cite{Shap83}):
\be
\lim_{x\rightarrow x_\rmscr{max}} \theta(x;y>1/3) = \infty.
\ee
where
\be
x_\rmscr{max} = 3\sqrt{3} y \sqrt{1-2y}.
\ee
 
\section{Light Curves}
\label{sec:lightc}

To construct a light curve, we must specify several angles ($\xi,
\kappa, \beta$ and $\gamma$).  During some portion of the precession
period, the observed flux will be constant with orbital phase, if
either $\beta=\kappa$ or $\xi=\kappa$.  In the first case, the angular
velocity vector will coincide with the location of the hotspot on the
star. In the second case, the angular velocity vector will point along
the line of sight once during each precession.

\subsection{$\beta=\kappa=90^\circ$}

To maximize the flux during the portion of the precessional period
where the flux does not vary with rotational phase, we take $\gamma=0$
and $\beta=\kappa=90^\circ$, and to minimize the flux during the rest of the
precession, we take $\xi=\theta(1)$.  If $\gamma \neq 0$, the maximum
flux will not occur during that portion of precession when the flux
does not vary with rotational phase.  Taking $\beta=\kappa \neq
90^\circ$ will reduce the maximum flux, and $\xi\neq\theta(1)$ will
change the portion of the time when the hotspot is not visible.

\figref{flux1} shows the observed flux from the hotspot as a function
of the star's precessional and rotational phase.  Only half of a
precessional period is depicted.  The flux varies at twice the
precession rate. As \figref{mflux1} shows, the mean flux over the
rotational period varies nearly sinusoidally.  Twice during the
precessional period, when the mean flux reaches its maximum, the flux
does not vary with the rotational phase of the star.   At other stages
of the star's precession, the hotspot spends much of the rotational
period hidden behind the horizon on the neutron star surface.  If
$\beta=\kappa \neq 90^\circ$, one finds that the mean flux varies at
the precessional frequency, resulting in a light curve similar to that
presented in \figref{flux2}.

\figcomment{
\begin{figure}
\plottwo{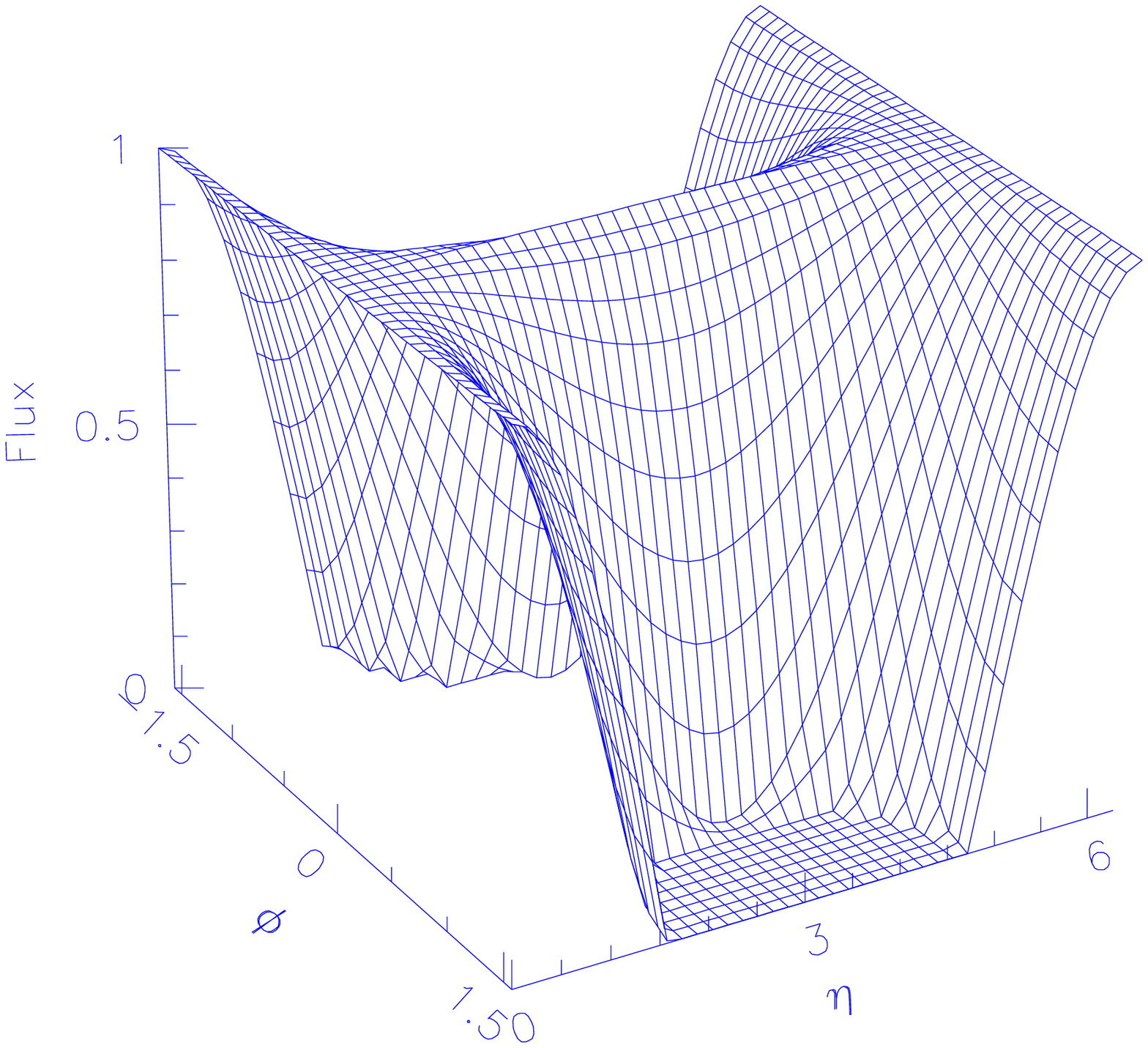}{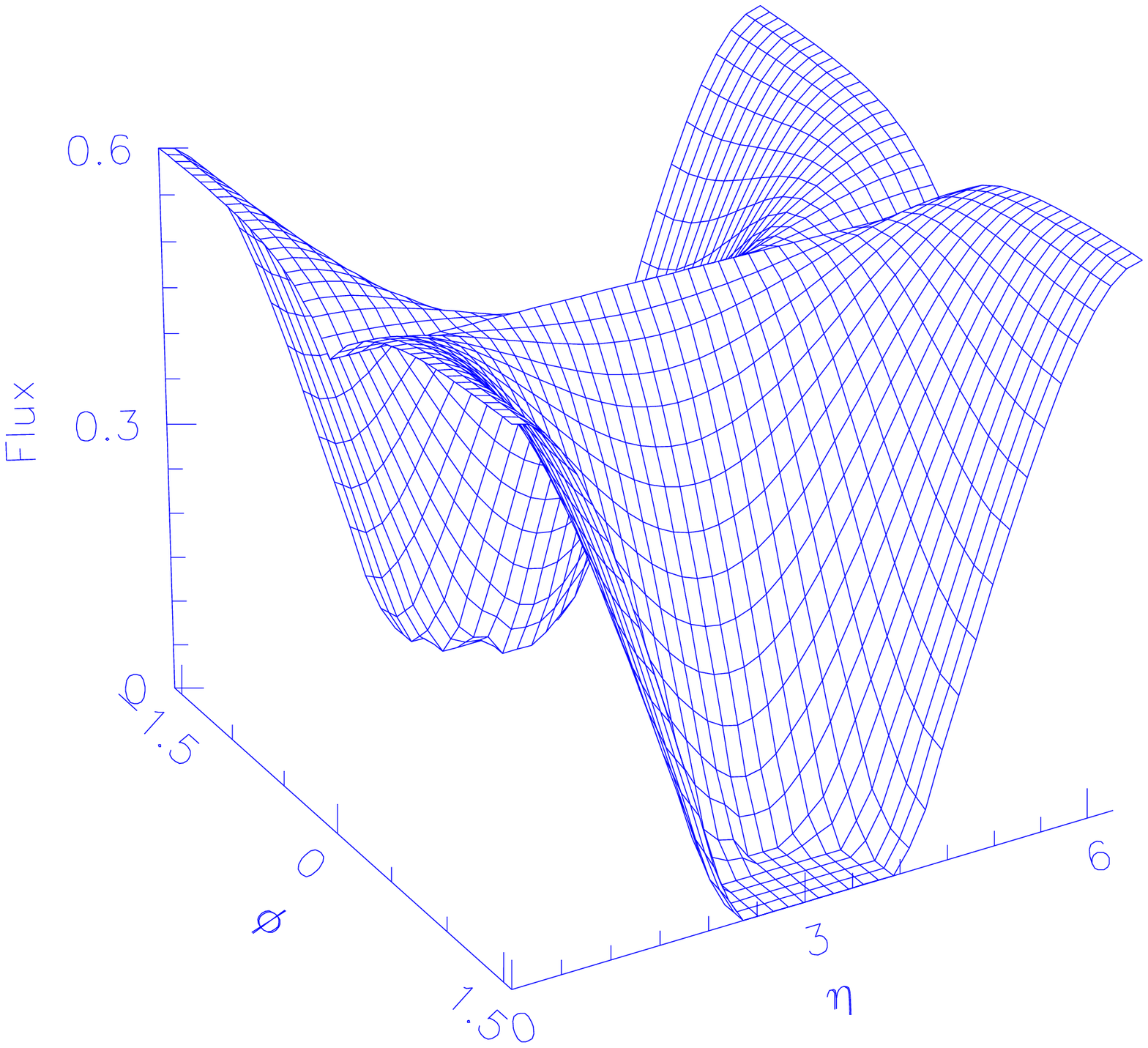}
\caption{Light curves as a function of precessional phase ($\phi$) and
rotational phase ($\eta$) for $y=0,0.2$ or $R=\infty$ and 11~km for a
1.4~M$_\odot$ neutron star.  For both models, $\kappa=\beta=90^\circ$,
$\gamma=0$ and $\xi=\theta(1)$.  A hotspot on a Newtonian star
directly below the observer has a flux of unity in these units.}
\label{fig:flux1}
\end{figure}
}

\figcomment{
\begin{figure}
\plottwo{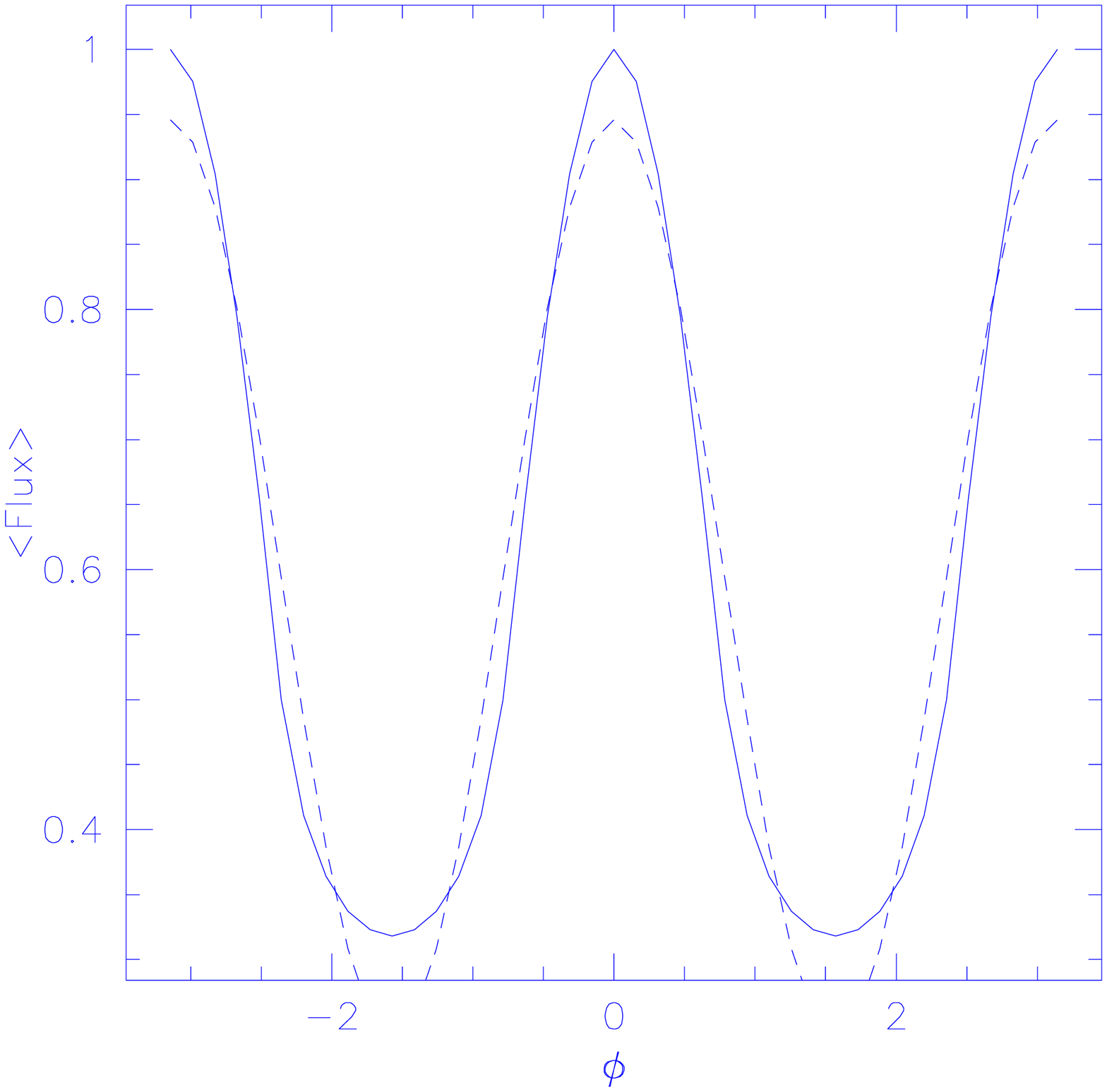}{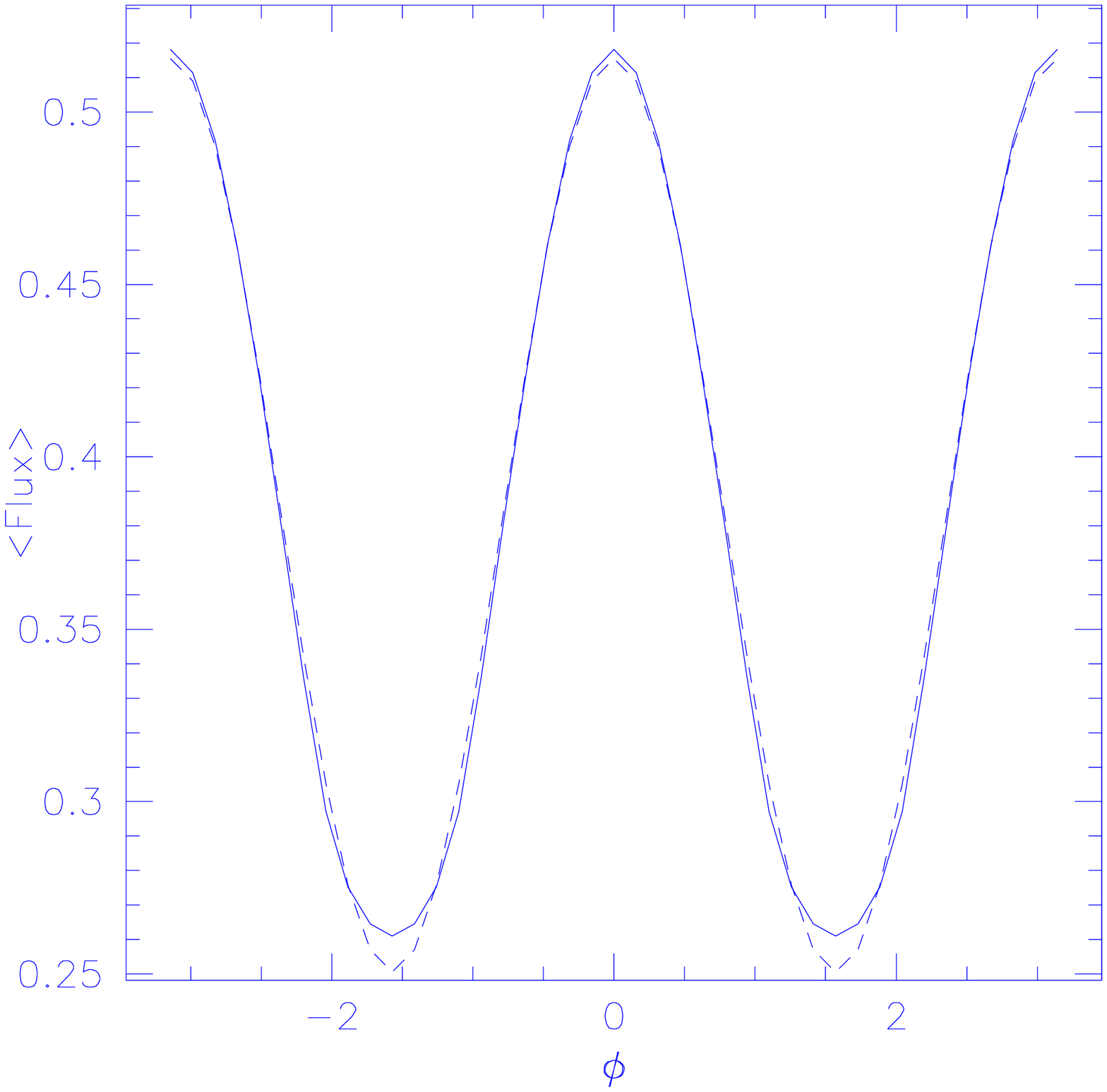}
\caption{Mean flux as a function of precessional phase ($\phi$)
for $y=0,0.2$ or $R=\infty$ and 11~km for a 1.4~
M$_\odot$ neutron star.
The dashed curve gives the best-fitting function of the form
$a\cos2\phi+b$. The geometry is as above.}
\label{fig:mflux1}
\end{figure}
}

\subsection{$\xi=\kappa=\theta(1)$}

In this case, we attempt to maximize the flux during the portion of
the precession when the flux does not vary over the rotation.  To do
this, we take $\beta=\xi=\kappa=\theta(1)$ and $\gamma=0$.  If the
bending of the photon trajectories is neglected (\ie as $y$ approaches
zero), $\theta(1)$ approaches $90^\circ$ and this case reduces to the
previous one.  However, for a realistic neutron star with $y\approx
0.2$, we have new light curve which varies at the precession rate (not
twice that rate as the previous case).  Taking $\beta\neq\kappa$ or
$\gamma\neq 0$ also yields a portion of the light curve when the flux
does not vary with rotational phase, but this does not coincide with
the period when the mean flux reachs its maximum.

\figcomment{
\begin{figure}
\plottwo{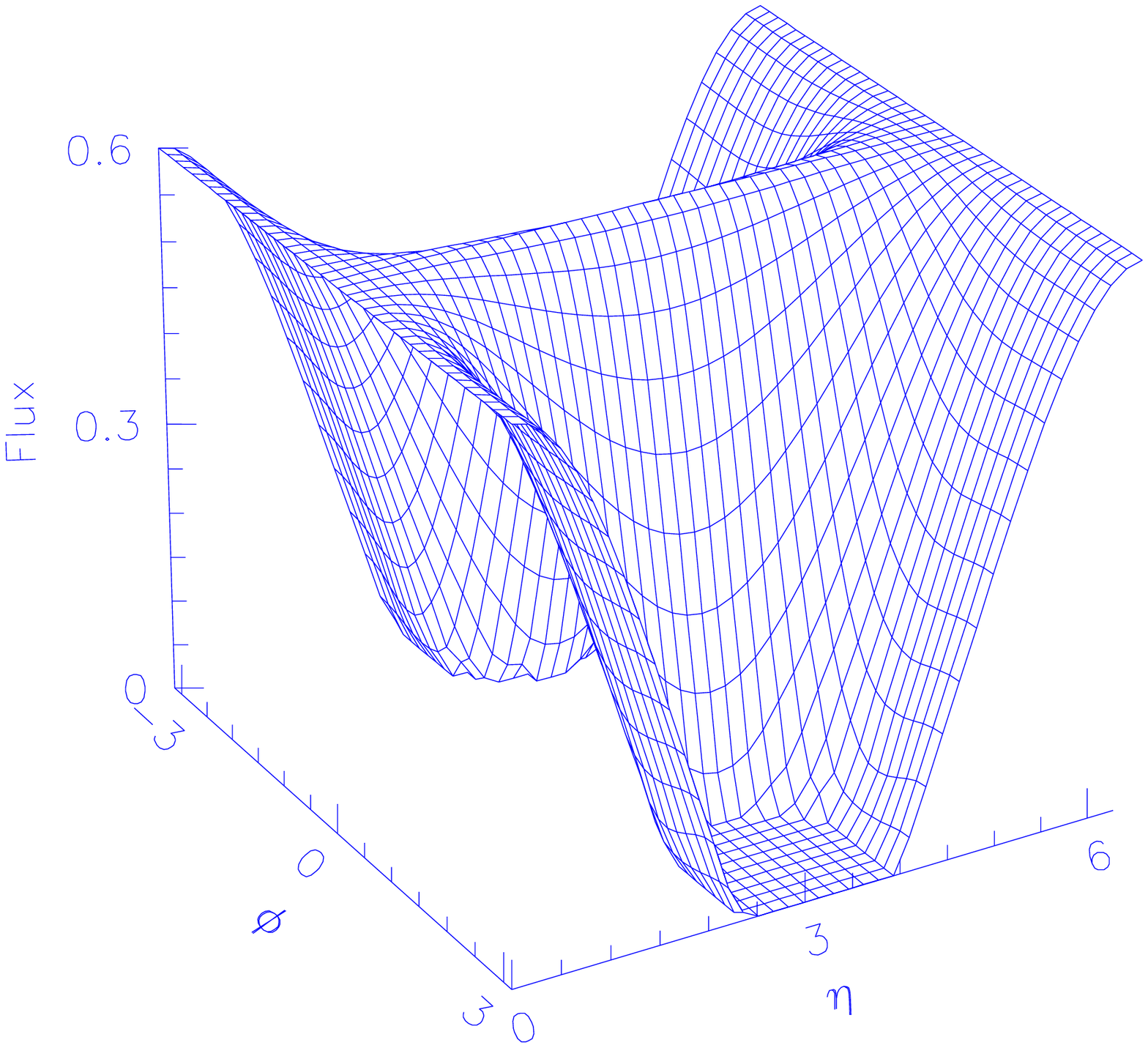}{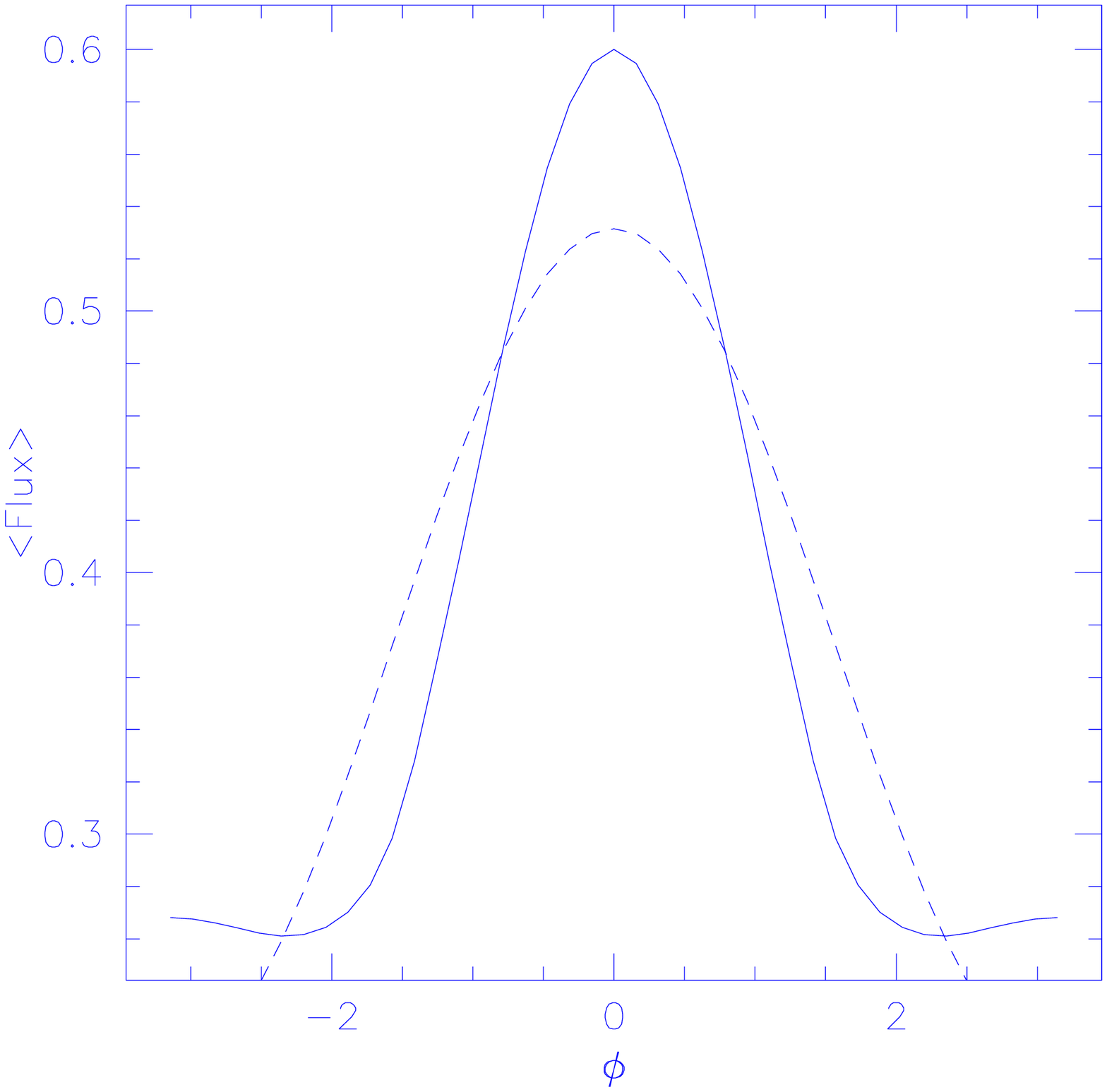}
\caption{The left panel gives the light curves as a function of
precessional phase ($\phi$) and rotational phase ($\eta$) for $y=0.2$
$R=11$~km for a 1.4~M$_\odot$ neutron star.
$\kappa=\beta=\xi=\theta(1)$ and $\gamma=0$ and $\xi=\theta(1)$.  The
right panel depict the mean flux as a function of $\phi$.  The dashed
curve gives the best-fitting function of the form $a\cos\phi+b$.}
\label{fig:flux2}
\end{figure}
}

\subsection{Random Geometry}

The choices of $\xi, \gamma, \beta$ and $\kappa$ that we have made
previously are not generic.  One would expect the values of $\xi$ and
$\gamma$ from a particular neutron star to be random.  The values of
$\beta$ and $\kappa$ are intrinsic to the star; therefore, one may
find a physical motivation for a particular distribution of their
values.  \figref{flux3} presents two light curves for two randomly
selected geometries.
\figcomment{
\begin{figure}
\plottwo{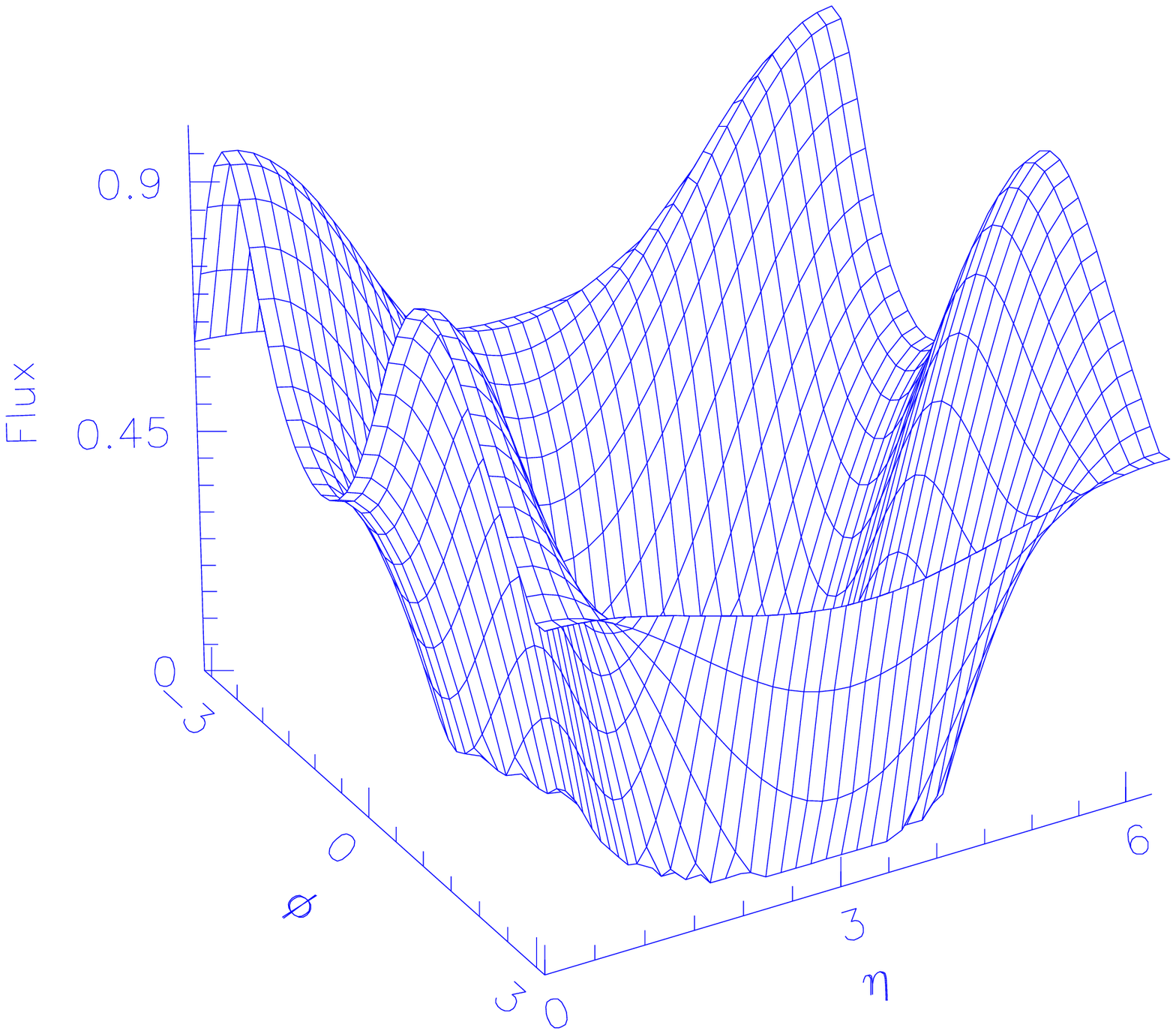}{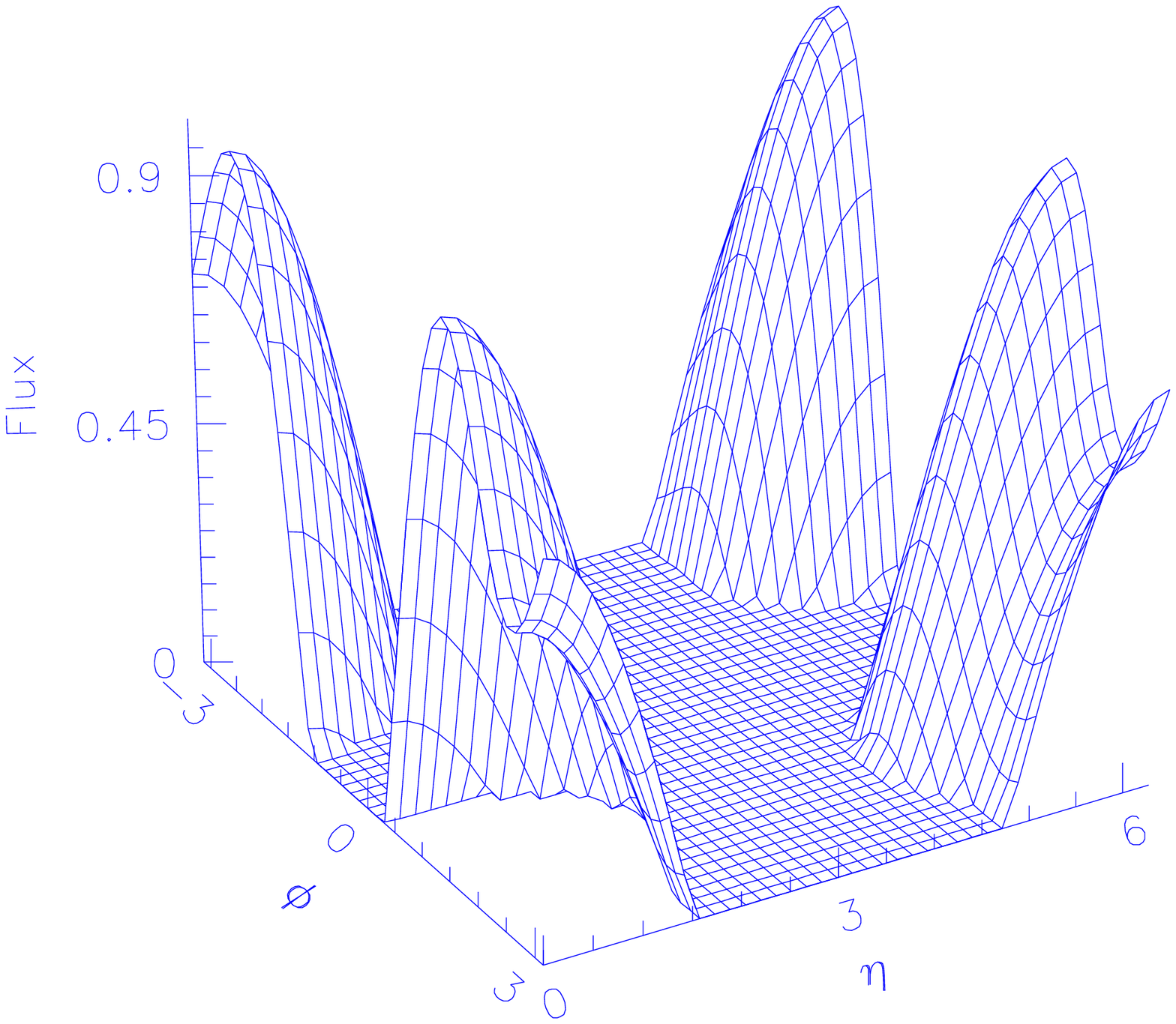}
\caption{The left panel gives the light curves as a function of
precessional phase ($\phi$) and rotational phase ($\eta$) for $y=0$.
The left panel has $\gamma=69^\circ, \kappa=144^\circ, \beta=62^\circ$
and $\xi=115^\circ$ and the right panel has $\gamma=138^\circ,
\kappa=96^\circ, \beta=126^\circ$ and $\xi=141^\circ$.}
\label{fig:flux3}
\end{figure}
}

We created a random sample of several hundred geometries and found that
approximately three percent yielded light curves qualitatively similar
to those presented in \figref{flux1} and \figref{flux2}.  Although one
would generally expect more complicated light curves 
such as those depicted in
\figref{flux3}, a significant fraction of the geometries yield
light curves in which the flux is constant with rotational phase when
it reaches its maximum, and during the rest of the precessional
period, the hotspot is hidden for a large portion of each rotation.

\section{Discussion} 
\label{sec:disc}

We have explored several possible light curves of a freely precessing
neutron star with a hotspot, and focussed on those whose flux is
constant with rotational phase when the flux reaches a maximum value.
These possibilities indicate that \rcwx\ may be a freely precessing
neutron star.  However, as the mean flux decreases from its maximum
it also begins to vary with the rotation of the star; therefore, in
the context of this model, we would expect that subsequent
observations of \rcwx\ may uncover its rotational period which we
would expect to be on the order of several seconds.  The variation of
the flux with the rotational phase of the star during some portion of
the precession appears generic to freely precessing stars with a
hotspot.

The Earth undergoes a free precession with a period of about 433~days,
known as the \jcite{1891AJ.....11...59C} wobble.  This is
significantly longer than one would expect from the asymmetry of the
Earth's figure due to dissipation inside the Earth (\cite{Burs93}).
Without excitation the wobble would disappear within about a century,
and the avenue for its excitation is still unclear.

In neutron stars, precession has been proposed to explain long-term
variations in their spin and pulse profiles (\eg
\cite{Davi70,Gold70,Rude70,Brec72,Pine72,Pine73,Pine74}).  If neutron
stars rotate as rigid bodies the precessional period would be
$P/\epsilon$ (\cite{Pine72,Pine74}) where $P$ is the rotational
period.  \jcite{1974ApJ...190..137R} proposed that neutron stars
contain a superfluid component in their cores.

\jcite{1977ApJ...214..251S} explored how the pinning of the superfluid
vortices affects the free precession of a neutron star.  He argued that
the dissipation timescale for the precessional mode ($\tau_w$) is of
the order of the postglitch relaxation time ($\tau$) times the ratio
of the rotational to the precessional frequency.  $\tau$ ranges from a
week for the Crab to nearly a century for 1641-45 (\cite{Shap83});
therefore, depending on the nature of the superfluid coupling the
precessional mode may last for millennia.  \jcite{1977ApJ...214..251S}
also found that if the star is triaxial the geometry of the precession
is more complicated that for the purely free precession considered
here.  Additionally, the precessional frequency in this case is given
by angular velocity of the superfluid component of the star times the
fractional contribution of the superfluid to the total moment of
inertia of the star (about one percent).

\jcite{1999ApJ...524..341S} have recently reexamined the precession of
multicomponent neutron stars with imperfect vortex pinning and found
several possibly long-lasting precessional modes with long periods
like the precession described here.  The excitation and decay of
precessional motions in neutron stars are still uncertain.

Evidence has been found for free precession in some radio pulsars
(\eg\ \cite{1997A&A...324.1005C,1988MNRAS.235..545J}), but it is not
generic (\eg\ \cite{1995AAS...187.1603M}); therefore, the question
arises as to which properties of a neutron star would allow or prevent it
from precessing and would they correlate with its radio emission.
\jcite{1999ApJ...519L..77M} and \jcite{2000MNRAS.313..217M} argue that
precession is characteristic of strongly magnetized neutron stars (\cf\
\eqref{bdistort}). \jcite{Usov96} and \jcite{Aron98} 
have proposed that
strongly magnetized neutron stars are unlikely to produce radio
emission collectively due to the formation of bound electron-positron
pairs.  Alternatively, \jcite{Bari97b} suggest that in sufficiently
strong fields ($B \gtrsim B_c$) the QED process of photon splitting
(\cite{Adle71}; \cite{Heyl97hesplit}) can dominate one-photon pair
production.  This will effectively quench the pair cascade, making
coherent pulsed radio emission impossible.

Since the timescales for both the excitation and decay of precessional
motion in neutron stars are unknown, one can appeal to the relative
youth of \rcwx\ and the other members of the AXP class.   They are
all several thousand years old, much younger than vast majority of
radio pulsars (\cite{Tayl93}).   The appropriate timescales for
precession may simply be shorter than the ages of most radio pulsars
while longer than those of AXPs.  Furthermore, the hints of precession
seen in the Crab pulsars (\cite{1997A&A...324.1005C}) point toward
this explanation.

We have examined the light curves of free precessing neutron stars
with a hotspot and focussed on those geometries which exhibit an epoch
during each precessional period where the flux does not vary as the
star rotates.   These geometries account for about three percent of a
random sample and may provide an explanation for the emission from
\rcwx\ .  If this is the case, further observations of the light curve
from \rcwx\ should reveal a pulse period of the order of $10^{-4}$
times the precessional period of six hours.  Free precession may be a
hallmark of young or highly magnetized neutron stars, and it is a
direct probe of the structure of the crust and interior of the neutron
star and the coupling between them.

\bibliographystyle{jer}
\bibliography{ns,physics,earth,precess,mine,qed}

\figcomment{
\end{document}
\end
}

\cleardoublepage

\section*{Figure Captions}

\begin{figure}
\caption{Definitions of various angles}
\label{fig:geom}
\end{figure}

\begin{figure}
\caption{Light curves as a function of precessional phase ($\phi$) and
rotational phase ($\eta$) for $y=0,0.2$ or $R=\infty$ and 11~km for a
1.4~M$_\odot$ neutron star.  For both models, $\kappa=\beta=90^\circ$,
$\gamma=0$ and $\xi=\theta(1)$.  A hotspot on a Newtonian star
directly below the observer has a flux of unity in these units.}
\label{fig:flux1}
\end{figure}

\begin{figure}
\caption{Mean flux as a function of precessional phase ($\phi$)
for $y=0,0.2$ or $R=\infty$ and 11~km for a 1.4~
M$_\odot$ neutron star.
The dashed curve gives the best-fitting function of the form
$a\cos2\phi+b$. The geometry is as above.}
\label{fig:mflux1}
\end{figure}

\begin{figure}
\caption{The left panel gives the light curves as a function of
precessional phase ($\phi$) and rotational phase ($\eta$) for $y=0.2$
$R=11$~km for a 1.4~M$_\odot$ neutron star.
$\kappa=\beta=\xi=\theta(1)$ and $\gamma=0$ and $\xi=\theta(1)$.  The
right panel depict the mean flux as a function of $\phi$.  The dashed
curve gives the best-fitting function of the form $a\cos\phi+b$.}
\label{fig:flux2}
\end{figure}

\begin{figure}
\caption{The left panel gives the light curves as a function of
precessional phase ($\phi$) and rotational phase ($\eta$) for $y=0$.
The left panel has $\gamma=69^\circ, \kappa=144^\circ, \beta=62^\circ$
and $\xi=115^\circ$ and the right panel has $\gamma=138^\circ,
\kappa=96^\circ, \beta=126^\circ$ and $\xi=141^\circ$.}
\end{figure}

\end{document}